\documentclass[10pt]{iopart}
\usepackage{epsfig}
\usepackage{iopams}
\usepackage{color}

\def\be{\begin{equation}}
\def\ee{\end{equation}}

\def\alt{\mathrel{\hbox{\rlap{\hbox{\lower4pt\hbox{$\sim$}}}\hbox{$<$}}}}
\def\agt{\mathrel{\hbox{\rlap{\hbox{\lower4pt\hbox{$\sim$}}}\hbox{$>$}}}}

\begin{document}

\title[Non-conservation of Carter in black hole
spacetimes]{Non-conservation of Carter in black hole spacetimes}

\author{Alexander Grant\dag, \'{E}anna \'{E}.\ Flanagan\dag}

\address{\dag\ Department of Physics, Cornell
University, Ithaca, NY 14853}

\date{\today}
\begin{abstract}

Freely falling point particles in the vicinity of Kerr black holes
are subject to a conservation law, that of their Carter constant.
We consider the conjecture that this conservation law is a special
case of a more general conservation law, valid for arbitrary processes
obeying local energy momentum conservation.  Under some fairly general
assumptions we prove that the conjecture is false: there is no
conservation law for conserved stress-energy tensors on the Kerr
background that reduces to conservation of Carter for a single point
particle.

\end{abstract}
\maketitle



The validity of conservation laws in physics can depend on
the type of interactions being considered.  For example, baryon number
minus lepton number, $B-L$,
is an exact conserved quantity in the Standard Model of particle
physics.  However, it is no longer conserved when one enlarges the set
of interactions under consideration to include those of many grand
unified theories \cite{Georgi:1974sy}, or those that occur in quantum
gravity \cite{Hawking1979175}.

The subject of this note is the conserved Carter constant for
freely falling point particles in black hole spacetimes
\cite{PhysRev.174.1559}.  As is well known, this quantity is not
associated with a spacetime symmetry or with Noether's theorem.
Instead it is obtained from a symmetric tensor $K_{ab}$ which obeys
the Killing
tensor equation $\nabla_{(a} K_{bc)} =0$ \cite{1970CMaPh..18..265W},
via $K = K_{ab} k^a k^b$, where $k^a$ is four-momentum.
Is this conservation law
specific to freely falling particles, or does it persist in the
presence of more general interactions?

The Carter constant conservation law has already been generalized in a
variety of directions:
\begin{itemize}
\item It applies to charged particles in
rotating charged black hole spacetimes \cite{PhysRev.174.1559}.

\item It has been generalized to spinning test particles, to linear
  order in the spin \cite{1983RSPSA.385..229R}
    \footnote{The motion of a spinning point particle in the Kerr
      spacetime is thus integrable to linear order in spin \cite{Hinderer:2013uwa}.
      This does
      not contradict the fact that chaotic behavior is seen in
      numerical studies of spinning point particle dynamics \cite{Kiuchi:2004bv}, since
      that behavior is due to effects that are higher order in spin.}.

\item For a free scalar field on the Kerr background, Carter showed
  that the differential operator ${\cal D} = K^{ab} \nabla_a \nabla_b
  +
(\nabla_a K^{ab}) \nabla_b$ commutes with the d'Alembertian, which
implies the existence of a conserved quantity for the field
\cite{PhysRevD.16.3395}.
For a free complex scalar field $\Phi$, the charge associated
with the conserved current
$
j^a = i \left[ ( {\cal D} \Phi)^* \nabla^a \Phi - (\nabla^a
  {\cal D} \Phi)^* \Phi\right]/2
$
is a generalization of the Carter constant for particle motion, in the
following sense.  Solutions of the form $\Phi \propto \exp[i
\varphi/\varepsilon]$ in the eikonal limit $\varepsilon \to 0$ can be
interpreted as streams of particles, and the conserved charge is just
the sum of the Carter constants of the particles.
This is valid for both massive and massless fields.

\item There is also a conserved quantity for spin $1/2$ fields related
  to the Killing tensor.
On the Kerr spacetime the Killing tensor can be expressed as the
square of an antisymmetric Killing-Yano tensor $f_{ab}$,
for which $\nabla_{(a} f_{b)c}=0$ and $K_{ab}=f_{ac}f_b{}^c$
\cite{NYAS:NYAS125,Thesis:Floyd}.
Carter and McLenaghan \cite{PhysRevD.19.1093} showed that the operator
$
i \gamma_5 \gamma^a (f_a^{\ b} \nabla_b - \gamma^b \gamma^c \nabla_c
 f_{ab}/6)
$
commutes with the Dirac operator and so gives rise to a conserved
quantity.
A similar construction for spin-1 fields can be found in Ref.\ \cite{castillo}.

\item Recently, Ashtekar and Kesavan have shown that in spacetimes
  which settle down at late times to a Kerr black hole, the Killing
  tensor at future null infinity can be expressed as a linear
  combination of products of asymptotic symmetry vector fields (BMS
  generators), allowing them to compute a charge associated with any
  cut and derive an asymptotic conservation law \cite{Kesavan}.

\item There is no known general local conservation law associated with the
  Killing tensor for spin 2 fields in Kerr, that is, for linearized
  vacuum perturbations.  However, there are hints that a conserved
  current may exist in this case.  Specifically, for the
  radiation-reaction inspirals of point particles into black holes, it
  is possible to compute the time averaged time derivative of the
  particle's Carter constant
  \cite{Mino:2003yg,2005CQGra..22S.801D,2005PThPh.114..509S,2006PThPh.115..873S}.
  The result consists of two terms, a term involving the amplitudes of
  the gravitational wave modes at future null infinity that is naturally
  interpreted as a flux to infinity, and a term involving the
  amplitudes of the modes at the horizon that is naturally interpreted
  as a flux down the horizon.

\end{itemize}

In this note we consider a different possible type of generalization
of the conservation law, the possibility that the Carter quantity
may be conserved under local interactions between particles that obey
local stress energy conservation.
Specifically, suppose we are given a conserved symmetric tensor
$T_{ab}$ on the Kerr spacetime, with compact spatial support.  Does
there exist a quantity ${\cal K}_\Sigma$ which can be computed from
$T_{ab}$ and its derivatives on any Cauchy surface $\Sigma$, which has
the properties that (i) ${\cal K}_\Sigma$ is independent of $\Sigma$,
and (ii) ${\cal K}_\Sigma$ reduces to the Carter constant for a single
point particle?  For example, one could consider quadratic functionals of
the form\footnote{The functional (\ref{ans}) is intended as an illustrative example only;
the argument below is not restricted to functionals of this form.}
\be
{\cal K}_\Sigma = \int_\Sigma d^3 \Sigma_a(x) \int_\Sigma d^3
\Sigma_{a'}(x')  T^{ab}(x) K_{bb'}(x,x') T^{a'b'}(x').
\label{ans}
\ee
Such functionals satisfy property (ii) if the bitensor $g_a^{\ b'}(x,x') K_{bb'}(x,x')$
reduces to the Killing tensor $K_{ab}$ in the coincidence limit $x'
\to x$ (see \ref{explicit}).  Of course, they do not necessarily
satisfy property (i).
If a conserved quantity of this type existed, one could say that the
Carter constant is conserved not just for freely falling,
non-interacting particles, but also under general processes that obey
local stress energy conservation.

\begin{figure}
\begin{center}
\epsfig{file=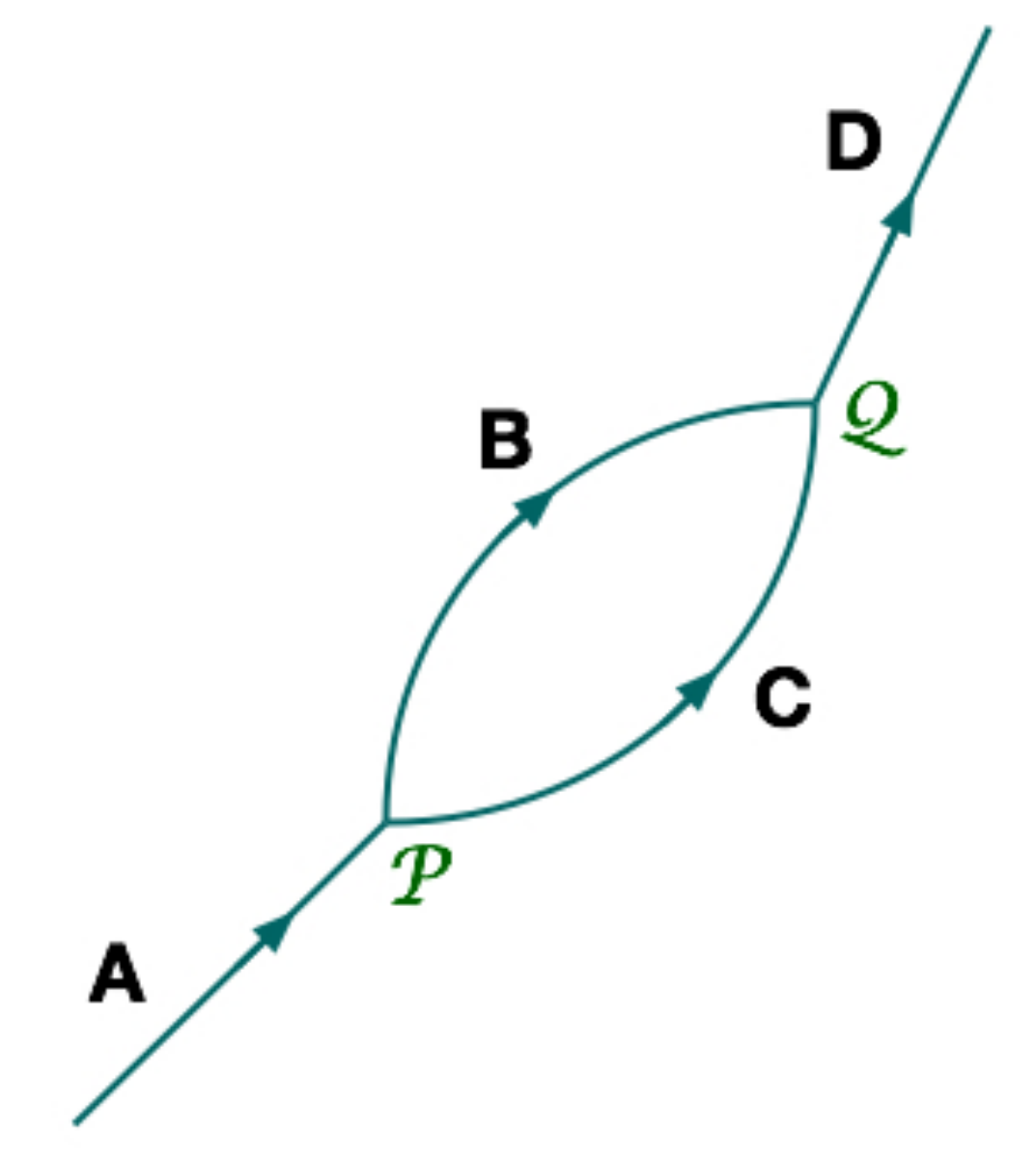,angle=0,width=5.0cm}
\caption{An illustration of a process in which local energy momentum
  conservation is satisfied but in which Carter constant conservation
  is violated.  A particle $A$ is freely falling in the Kerr
  spacetime.  At some point ${\cal P}$ it splits into two particles
  $B$ and $C$.  Those particles then freely fall, and collide at some
  later event ${\cal Q}$ to form a single particle $D$.  The Carter
  constants of $A$ and $D$ do not coincide.}
\label{fig:modes}
\end{center}
\end{figure}

No such conservation law exists.  There is a very simple argument
which shows this, which is as follows.  Consider a freely falling
particle $A$ in the Kerr spacetime.  Suppose that at some point ${\cal
  P}$, it splits into two particles $B$ and $C$, obeying
energy-momentum conservation.  Thus the four-momenta ${\vec k}$ of the
particles are
related by
\be
{\vec k}_A({\cal P}) = {\vec k}_B({\cal P}) + {\vec k}_C({\cal P}).
\ee
Suppose that the initial four-momenta and masses of $B$ and $C$ are carefully chosen
so that these two particles, after falling freely for some time, come
together again at some other
spacetime point ${\cal Q}$.  At that point the two particles combine
to form a fourth particle, $D$, again obeying energy-momentum
conservation:
\be
{\vec k}_D({\cal Q}) = {\vec k}_B({\cal Q}) + {\vec k}_C({\cal Q}).
\ee
Between the collisions, the Carter constants of the individual particles
$
K_A = K_{ab} k^a_A k^b_A,
$
etc, of the individual particles are locally conserved.

Now, if a general conservation law of the type discussed above existed,
then at early times the putative conserved quantity ${\cal K}$ would
be simply
the Carter constant of $A$, i.e
\begin{eqnarray}
{\cal K} &=& K_A = K_{ab}({\cal P}) k^a_A({\cal P}) k^b_A({\cal P})
\nonumber \\ &=&
K_B + K_C + 2 K_{ab}({\cal P}) k^a_B({\cal P}) k^b_C({\cal P}). \ \ \
\label{eq:initial}
\end{eqnarray}
Similarly at late times\footnote{At intermediate times, between the two collisions, the value of the
conserved quantity ${\cal K}$ would depend on the details of how the
conserved quantity is supposed to be computed.  For example, in the
model (\ref{ans}) it would depend on values of the bitensor $K_{ab'}$
away from the coincidence limit.  This is why we need to consider a process that begins and ends with a single particle, in order to get a clean test of the conjecture.} it would be
\begin{eqnarray}
{\cal K} &=& K_D = K_{ab}({\cal Q}) k^a_D({\cal Q}) k^b_D({\cal Q})
\nonumber \\
&=&K_B + K_C + 2 K_{ab}({\cal Q}) k^a_B({\cal Q}) k^b_C({\cal Q}). \ \ \
\label{eq:final}
\end{eqnarray}
However, by numerically integrating the geodesic equations in Kerr, it
is easy to find examples of scenarios of this kind where the initial
and final values (\ref{eq:initial}) and (\ref{eq:final}) do {\it not}
coincide (see \ref{numerical}).  This occurs whenever the cross term
$K_{ab} k^a_B k^b_C$ has different values at ${\cal P}$ and ${\cal Q}$.

In conclusion, generalizations of the Carter constant conservation law
are relevant to the program of computing gravitational wave signals
from point particles inspiralling into spinning black
holes\cite{Detweiler_rev05,MPP,Barack_rev,Barack_rev14}.
One might have hoped for a unified conservation law for ``Carter'',
analogous to that for energy, that some combination of the Carter
constant of the particle and the ``Carter'' in the gravitational wave
field be conserved.  The result presented here suggests that
no such unified law exists.

\ack
This research was supported in part by NSF grants PHY-1404105
and PHY-1068541.  We thank
Scott Hughes, Justin Vines, David Nichols and Leo Stein for helpful conversations.

\appendix

\section{Numerical Example}
\label{numerical}

We use Boyer-Lindquist coordinates $(t,r,\theta,\phi)$, choose units
so that the mass of the black hole is $M = 1$, and choose the value of
the spin parameter to be $a = 0.84$.  The event ${\cal P}$ is
$(0,3.508712731567,\pi/2,0)$, and ${\cal Q}$ is
$(48.77745940108,3.533582213837,1.549703133614,7.832144679602)$.
The contravariant components of $B$'s four-velocity at ${\cal P}$ are
$$
{\vec u}_B({\cal
  P})=(2.043684586293,0.00004124671403569,0.2549376720024,0.1436563472919),
$$
while those of $C$ are
$$
{\vec u}_C({\cal P})= (2.083715919305,0.01417859756406,-0.2712642321094,
0.1252970976976).
$$
These four-velocities were chosen to be close to the case where
$u_C^\theta = -u_B^\theta$ and re-intersection is guaranteed
(although one can show that $K_A = K_D$ in this special case).  We choose the masses of $B$ and
$C$ to be unity, so that $\vec{k}_B \cdot
  \vec{k}_B = \vec{k}_C \cdot \vec{k}_C = -1$.  The interval of
proper time from ${\cal P}$ to ${\cal Q}$ for $B$ is $\Delta \tau_B =
20.424435554150$,
while for $C$ it is $\Delta \tau_C = 24.10742349604$.  Our numerical
integration code conserves the quantities $E$, $L_z$, and $K$ along
each geodesic to within one part in $10^{12}$, and each geodesic
reaches the event $\mathcal{Q}$ to within one part in $10^{12}$ as well.
With these choices the
initial Carter constant is $K_A = 0.04154646396564$, while the final one is $K_D = 0.04392127426890
$.
The fractional difference is
$$
\frac{2(K_A - K_D)}{K_A  + K_D} = -0.055572087 \pm 0.000000002,
$$
which is nonzero.

We have also constructed several other numerical examples.  Those examples are sufficient to prove
a slight generalization of the result in the body of the paper: there
is no general conservation law for conserved stress-energy tensors
associated with any Killing tensor of the form $K_{ab} + \lambda g_{ab}$ for any value of $\lambda$.

\section{Observables that reduce to Carter constant for point
  particles}
\label{explicit}

In this appendix we show that the expression (\ref{ans})
reduces to the Carter constant of a point particle when the stress energy tensor is taken
to be that of a point particle.  If the particle's worldline is
written as $x^\alpha = z^\alpha(\lambda)$, then this stress energy
tensor is
\be
T^{\alpha\beta}(x) = \int d\lambda \, p^\alpha (\lambda) p^\beta (\lambda)
\frac{\delta^{(4)}(x - z(\lambda))}{\sqrt{-g}},
\ee
where $p^\alpha = dz^\alpha / d\lambda$.  We now insert this into the expression (\ref{ans}).
We use Gaussian normal coordinates adapted to the surface $\Sigma$, so that the surface is $t=t_0$, the metric is
$ds^2 = - dt^2 + h_{ij} dx^i dx^j$, and $d^3 \Sigma_a = \sqrt{h} n_a d^3x$ with ${\vec n} = \partial_t$ the unit normal.
The result is
\begin{eqnarray}
\fl
{\cal K}_\Sigma = \int dt\,  d^3 x \sqrt{h} n_\alpha \,
\int dt' \, d^3x'    \sqrt{h'} n_{\alpha'}
\int d\lambda \int d\lambda' K_{\beta\beta'}[z(\lambda),z(\lambda')]
\nonumber \\
\fl
\ \ \ \  \times p^\alpha(\lambda) p^\beta(\lambda) p^{\alpha'}(\lambda') p^{\beta'}(\lambda') \delta[t - t(\lambda)] \delta[t' - t(\lambda')]
\frac{\delta^3[{\bf x} - {\bf z}(\lambda)]}{\sqrt{h}} \frac{\delta^3[{\bf x}' - {\bf z}(\lambda')]}{\sqrt{h'}}.
\label{expr1}
\end{eqnarray}
Evaluating the time integrals of the delta functions give factors of $|dt / d\lambda| = (n_\alpha p^\alpha)^{-1}$,
and so ${\cal K}_\Sigma$ reduces to $p^\beta p^{\gamma} K_{\beta\gamma}$ evaluated at the point where the worldline crosses the surface.
Since we are assuming that the bitensor reduces to the Killing tensor in the coincidence limit, this is just the Carter constant.

\section*{References}


\end{document}